\documentclass[epj]{svjour}
\usepackage{graphics}
\usepackage{epsfig}
\usepackage{amsmath}

\def\gsim{\,\lower.25ex\hbox{$\scriptstyle\sim$}\kern-1.30ex%
\raise 0.55ex\hbox{$\scriptstyle >$}\,}
\def\lsim{\,\lower.25ex\hbox{$\scriptstyle\sim$}\kern-1.30ex%
\raise 0.55ex\hbox{$\scriptstyle <$}\,}

\def\Journal#1#2#3#4{{#1} {\bf #2} (#3) #4}

\def\NPB{{\em Nucl. Phys.}   {\bf B}}
\def\PLB{{\em Phys. Lett.}   {\bf B}}

\def\PRD{{\em Phys. Rev.}    {\bf D}}

\def\ZPC{{\em Z. Phys.}      {\bf C}}

\def\EJC{{\em Eur. Phys. J.} {\bf C}}

\def\p{\mathrm{p}}
\def\n{\mathrm{n}}
\def\U{\mathrm{u}}
\def\u{\mathbf{u}}
\def\D{\mathrm{d}}
\def\d{\mathbf{d}}
\def\e{\mathrm{e}}
\def\X{\mathrm{X}}
\newcommand{\youngtab}[3]
{
\rlap{\raisebox{0.3\intextsep}{\framebox[14pt]{$\vphantom{U}#1$}}}
\raisebox{-0.3\intextsep}{\framebox[14pt]{$\vphantom{U}#2$}}
\raisebox{0.3\intextsep}{\framebox[14pt]{$\vphantom{U}#3$}}
}

\begin{document}
\title{The qq$\bar{\mathbf{q}}$ component in the constituent quark  generates the pion cloud of the nucleon}
\author{B.~Povh\inst{1} 
\thanks{\emph{e-mail:} b.povh@mpi-hd.mpg.de} 
   and M.~Rosina\inst{2}\inst{3}
\thanks{\emph{e-mail:} mitja.rosina@ijs.si}  
}                     
\institute{Max-Planck-Institut f\"{u}r Kernphysik,
           Postfach 103980, D-69029 Heidelberg, Germany
 \and Faculty of Mathematics and Physics, University of Ljubljana, Ljubljana, Slovenia,
  \and Jo\v{z}ef Stefan Institute, Jamova 39, P.O. Box 3000, SI-1001 Ljubljana, Slovenia
}
\date{Received: date / Revised version: date}
%
\abstract{The spin properties and the flavor asymmetric sea of the nucleon can be well explained
by assuming that each constituent quark is surrounded with about 30\% probability by a quark-antiquark pair coupled to the pion  quantum numbers.
We show that about one quarter of these quark-antiquark pairs show up as the pion in
$\p\to\n\pi^+$ and $\p\to\p\pi^0$ fluctuation in agreement with the observed value in the ($\e+\p\to\e+$ forward neutron+X) experiment.
\PACS{
      {12.38.-t}{Quantum chromodynamics}   \and
      {12.39.Jh}{Nonrelativistic quark model} \and
      {14.20.Dh}{Protons and neutrons}
     } 
} 
\maketitle
%
\section{Introduction}
The notion of the constituent quark applies generally to the massive quark dressed by gluons, the
constituent of the nucleon. This non-relativistic model with three massive constituent quarks 
 works well for the hadronic masses
and the magnetic moments. It breaks down if the spin properties of the baryons are considered.
The improved version, the chiral constituent quark model is surprisingly succesful in explaining the 
spin properties of nucleons and hyperons. In the simplest form applied to the nucleon the chiral constituent 
quark is 
composed of a massive quark accompanied by a quark-antiquark pair coupled to the spin-parity
quantum numbers of the pion $J^{\pi}=0^-$. In the following we write the pion symbol as a shortcut 
to the quark-antiquark pair coupled to the pion quantum numbers. This simple model has been
first applied by  Eichten et al. \cite{eichten} to explain the flavor asymmetry of the sea quarks
and further elaborated by Baumg\"artner et al.~\cite{BPKP} and Pirner~\cite{Pirner2} 
in the interpretation of the spin properties of the nucleon. 
It is related to the three-flavour extension proposed by Cheng and Li~\cite{Cheng}.
Explicitly written, the chiral constituent up-quark (u) structure is
\begin{equation}
|\U\rangle=\sqrt{(1-\frac{3}{2}\,a)}\,|\u\rangle+\sqrt{a}|\d\pi^+\rangle
+\sqrt{\frac{a}{2}}\,|\u\pi^0\rangle,
\label{eq:quark1}
\end{equation}
and of the down quark (d)  
\begin{equation}
|\D\rangle=\sqrt{(1-\frac{3}{2}\,a)\,}|\d\rangle+\sqrt{a\,}|\u\pi^-\rangle
+\sqrt{\frac{a}{2}}\,|\d\pi^0\rangle.
\label{eq:quark2}
\end{equation}
The basis of pure flavour quarks is denoted by boldface $\u$ and $\d$.

At Q$^2 \approx 0$ gluons do not appear as an explicit degree of freedom 
and the nucleon is composed of quarks and quark-antiquark pairs. 
Thus in the lowest order the Fock state of the constituent quark has the form  
(\ref{eq:quark1} and \ref{eq:quark2}),
 where in the second and third term the quark-antiquark pair 
is coupled to the $J^{\pi}=0^-$ quantum nummbers of the pion. 
This simple structure of the chiral constituent quark (\ref{eq:quark1}) 
has two attractive features. Firstly,
as we will show, the chiral constituent quark reproduces the experimental results of the 
deep inelastic scattering and axial-vector beta decays of the neutron quantitatively; secondly, 
this model complies with our 
picture of the origin of the quark mass by the chiral symmetry breaking mechanism of 
Nambu and Jona-Lasino
\cite{NJL}. Dressing the light quark by gluons is inevitably accompanied by creation of the Goldstone
boson, the pion. The Goldstone pion is an inherent part of the constituent quark.  

The parameter $a$ of (\ref{eq:quark1}, \ref{eq:quark2}) is usually determined from the value of the 
axial vector coupling constant $g_A=1.269\pm0.003$ \cite{PDG} yielding  $a=0.239\pm0.002$. 
The parameter $a$ measures the probability 
of the constituent quark to be in the state accompanied with a charged pion. Furthermore, 
with the probability $a/2$ the constituent quark is in a state component with the neutral pion.
Thus the total probability of finding a pion in the constituent quark amounts thus to
slightly more than one third. 
The large probability of the pion in the constituent quark is best manifested in the 
measurements of the quark polarization in the deep inelastic scattering. Not only
that one third of the constituent quark with the pion does not contribute to the
spin polarization, but even more, with the oppositely oriented quarks reduces 
the total quark polarization to one third of what would be without the pions.
The loss of the angular momentum because of the oppositely oriented quark
is compensated by the orbital angular momentum of the pion in the $p$-state.   
The comparison of the experimental results of the deep inelastic scattering with 
the prediction of the chiral constituent-quark model is given in \cite{povh}.
It is also worthwhile to mention that the valence-quark distribution does not
peak at Bjorken $x=0.3$ but it is softer and peaks at $x=0.2$ corresponding to five and not three 
constitunts of the proton even before gluons can get excited. Eichten et al. (\cite{eichten})
ascribe these quark-antiquark pairs to an asymmetric sea.

We consider also other observables which depend strong\-ly on the pions in the nucleon:
the Gottfried sum rule $I_G$ (with corrections discussed in \cite{povh}), 
the integrals of the spin structure functions of proton $I_\p$ and deuteron $I_\d$ 
and the quark spin polarization $\Delta\Sigma$ \cite{rith,Airapetian}. 
They have larger error bars than $g_A$, 
but the values of $a$ fitted to them are within one standard deviation with each 
other (Table 1.). The new experimental value for $\Delta\Sigma$ 
supports even more our assumption that the main contribution 
to the spin reduction comes from the pion fluctuation.
\vspace{10mm}
\begin{table}[hhh]
\vbox{\baselineskip 15pt\halign{
\hfil$#$\hfil\hskip5mm&$#$\hfil\hskip5mm&$#$\cr
\hfil\mathrm{observable}\hfil&\hfil \mathrm{model\, value} \hfil& \hfil a \hfil \cr
\noalign{\vskip 2mm\hrule\vskip 2mm}
g_A=1.269\pm0.003 & \frac{5}{3}(1-a)  & a=0.239\pm0.002 \cr
I_G=0.216\pm0.033 & \frac{1}{3}(1-2a) & a=0.176\pm0.050 \cr
I_\p=0.120\pm0.017& \frac{5}{18}(1-2a)& a=0.284\pm0.031 \cr
I_\d=0.043\pm0.006& \frac{5}{36}(1-3a)& a=0.231\pm0.014 \cr
\Delta\Sigma=0.330\pm0.064& (1-3a)    & a=0.223\pm0.021 \cr
\noalign{\vskip 2mm\hrule}}}
\caption{The $\pi^+$ probability $a$ fitted to different observables}
\end{table}

A rather large presence of the quark-antiquark pairs with the pion quantum numbers
as parts of the constituent quarks in the nucleon
ask for the answer of how do they show up in the low energy nuclear interaction.  
This answer is given below.
%
%
\section{Neutron plus pion is a Fock component of the proton}

Let us consider the matrix element $\langle \n\pi^+|p\rangle$.

Inserting for constituent quarks our chiral quarks it is evident that the $\langle \n\pi^+|$ 
has an overlapp with a Fock component of the proton. The result of the explicit calculation is 
\begin{equation}
|\langle \n\pi^+|\p\rangle|^2 =|\langle \D\pi^+|\U\rangle|^2 = (1-\frac{3}{2}\,a)\,a=0.15.
\label{eq:quark3}
\end {equation}
The result (\ref{eq:quark3}) means that the constituent u quark has a component 
of the d quark and a pion. It is however surprising that there is a missing 
factor of two as the proton has two u quarks. The reason for this is the 
flavor-spin-color structure of the nucleon. The flavor-spin wavefunction of the 
proton has a mixed symmetry combined into a symmetric flavor-spin function:
\begin{equation}
|\p\rangle 
= \sqrt{\frac{1}{2}}\quad{\youngtab{1}{3}{2}}_f\, \times \, {\youngtab{1}{3}{2}}_s \,
+ \sqrt{\frac{1}{2}}\quad{\youngtab{1}{2}{3}}_f\, \times\, {\youngtab{1}{2}{3}}_s \,.
\label{eq:quark4}
\end{equation}
A similar expression stays for the neutron. Since the combined wavefunction 
is symmetric under all permutations it is enough to look at the contribution of the particles 1 and 2.
In the first term of the proton wavefunction the particles 1 and 2 are symmetric 
and can both be u quarks and contribute constructively to the matrix element 
with a factor of two. In the second term the interference is destructive 
and the contribution cancels. Thus only the first term contributes to the matrix element.
Since both in proton and in neutron the first term appears 
with a factor $\sqrt{1/2}$, the factor two is canceled out.
This qualitative explanation can be verified by writing down the three-quark wavefunctions
explicitly.

This can be seen even  easier in the isospin formalism. In the act of producing a positive pion, 
the corresponding u quark loses one unit of charge, it becomes a d quark. 
This can be described with the operator $\sum_i t_-(i)=T_-$ where $T_-=T_x-\mathrm{i}T_y$.
We conveniently took the sum over all three quarks since the third quark, D, contributes zero anyway.
The expectation value is $ <TM-1|T_-|TM> = \sqrt{T(T+1)-M(M-1)}$ which for proton ($T=1/2, M=1/2$) 
gives in fact the factor 1. It is instructive to compare with $\Delta^+$ ($T=3/2, M=1/2$) 
in the process ep$\to e\Delta X$ where one  gets the factor 2, pointing 
out that the two u quarks are always symmetric and interfere constructively. 
Of course, for the squared amplitude, we get the additional factor
$a$ since only the $\pi^+$-dressed component of the u-quark contributes, and the factor $(1-\frac{3}{2}a)$ 
for the naked component of the final d-quark.
\section{Experimental test of the pion fluctuation}

The pion fluctuation of nucleon is well known in the classical nuclear
physics as anomalously large pion-nucleon coupling constant $g^2/4\pi=13.6$. Many of the nucleon
properties are ascribed to the pion cloud of the nucleon \cite{pioncloud}. 
Hovewer, there is no direct way of determinig experimentally
the probability of finding a pion fluctuation in the proton. The best way is to 
calculate the pion flow by using the pion-nucleon coupling constant and the form factor 
assuming that the pion is emitted by a proton \cite{PovhKop}, \cite{Holtmann} 
\begin{equation}
f_{\pi^+/\p}(x_L,t)=\frac{1}{2\pi}\frac{g^2_{\p\pi\n}}{4\pi}
(1-x_L)^{1-2\alpha(t)}\frac{-t}{(m_\pi^2-t)^2}|G(t)|^2.
\end{equation}
The pion flow is related to the measured cross section by
 \begin{equation}
\d\sigma^{\gamma^*\p\to \n\X}=f_{\pi^+/\p}(x_L,t)\cdot \d\sigma^{\gamma^*\pi^+\to \X}
\end{equation}
where the ($\gamma^*\pi^+\to \X$) DIS cross section is assumed to be 2/3 of the 
($\gamma^*\p \to \X$) DIS cross section in the cited analysis, 
with corrections due to absorption \cite{povh}.

Obviously the pion is not emitted by a proton but by a quark.
But as we showed above the state of the pion is dictated by the proton wave function
and the pion form factor simulated well the assumption that the emission
is from the proton. In the series of experiments \cite{H1LN}-\cite{H1LNjets}
measuring the spectrum of the forward neutrons in the reaction (e+p$\to$e+forward n+X)
has been shown that the high energy end of the neutron spectrum is consistent 
with the assumption that the deep inelastic scattering takes place on the pion.
Thus we are justified to say that the forward neutron is the signature of
the reaction taking place on the pion and that the total probability of finding 
a pion in ep$\to$n$\pi^+$ fluctuation can be obtained by integrating over the 
variables of the pion flow.

The analysis depends to some extent on the estimation of pion flux $f_{\pi^+/\p}$.
The analysis has been elaborated in \cite{povh} and the quoted results are
$\langle n\pi^+|p\rangle^2=0.165\pm0.01$ and $0.175\pm0.01$, respectively, 
for the two form factors best fitting to the experiment in \cite{PovhKop} and \cite{Holtmann}.

\section{Conclusion}
The pion fluctuation p$\to$n+$\pi^+$ and p$\to$p+$\pi^0$ is an artifact of the quark-antiquark pairs
of the constituent quarks. The impressive agreement between the measured and the calculated ratios
between the probability of the pion fluctuation and the probability of finding a quark-antiquark pair 
of the constituent quark  
is  a strong support of the constituent quark model. 

In this section we stress the difference between the notion of the
quark-antiquark pairs coupled to the pion quantum numbers being part of the constituent quarks
and the pions of the proton. While the quark-antiquark pairs are implied by the 
experimental values of $g_A$, the integrated spin structure functions and the violation 
of the Gottfried summ rule, the fluctuating pions  
are identified by the characteristic energy and $p_T$
distribution of the neutron spectra in the $\e\p\to\n\pi^+$ reaction.

Eichten et al. \cite{eichten} have named the quark-antiquark pairs of the constituent quark
the asymmetric quark sea. This name emphasizes hopefully sufficiently the difference
of their origin as compared to the normal quark sea. 

For the value $a=\langle \d\pi^+|\U\rangle^2=0.24$ each quark contains 0.36 quark-antiquark
pairs. Summing up the quark-antiquark pairs one obtains 
about one quark-antiquark pair per nucleon. Using this value of $a$ gives $\langle n\pi^+|p\rangle^2=0.15$.
This number 
corresponds well with the experimental value of $\langle n\pi^+|p\rangle^2=0.165\pm0.01$ or $0.175\pm0.01$.
It follows that in $\approx 0.26$ cases the proton is a neutron+ $\pi^+$ or a proton+$\pi^0$.
This means that about one quater of the nucleon's quark-antiquark pairs show up as
the pion fluctuation. 
\section*{Acknowledgments} We wish to thank A.~Bunyatyan for 
the discussions of the analysis on the forward neutron spectra and K.~Rith
for pointing us out the new results on spin-polarization data in DIS.


\begin{thebibliography}{99}

\bibitem{eichten}
E.J.Eichten, I.Hinchliffe and C.Quigg, \Journal{\PRD}{45}{1992}{2269}.

\bibitem{BPKP}
S.~Baumg\"artner, H.J.~Pirner, K.~K\"onigsmann and B.~Povh,
\Journal{ZP}{A 353}{1996}{397}.

\bibitem{Pirner2}
H.J.~Pirner, \it Prog.~Part.~Nucl.~Phys.\rm {\bf 36}~(1996)~19.

\bibitem{Cheng} T. P. Cheng and Ling-Fong Li,
   \it{ Phys. Rev. Lett.} {\bf 74} (1995) 2872-2875.

\bibitem{NJL}
S.P.~Klevansky, \it Rev.~Mod.~Phys. \rm {\bf 64}~(1992)~694.

\bibitem{PDG}
K.~Nakamura et al. (Particle Data Group), \it J.Phys. G \rm {\bf 37}~(2010)~075021.

\bibitem{povh}
A.~Bunyatyan and B.~Povh, 
\Journal{EJA}{A27}{2006}{359}.


\bibitem{rith}
K.~Rith,  \it Prog.~Part.~Nucl.~Phys.\rm {\bf 49}~(2002)~245.

\bibitem{Airapetian}
A.~Airapetian et al., 
\Journal{PRD}{75}{2007}{012007}.



\bibitem{pioncloud}
A.~W.~Thomas, Prog. Theor. Phys. {\bf 168} (2007) 614

\bibitem{PovhKop}
B.~Kopeliovich, B.~Povh and I.~Potashnikova, \Journal{\ZPC}{73}{1996}{125},
  [hep-ph/9601291].

\bibitem{Holtmann}
H.~Holtmann {\it et al.}, \Journal{\PLB}{338}{1994}{363}.


\bibitem{H1LN}
C. Adloff {\it et al.}  [H1 Collaboration], \Journal{\EJC}{6}{1999}{587},
   [hep-ex/9811013].

\bibitem{zeuslni} S.~Chekanov {\it et al.} [ZEUS Collaboration], 
\Journal{\NPB}{637}{2002}{3},
[hep-ex/0205076].

\bibitem{zeuslntraj}
S. Chekanov {\it et al.} [ZEUS Collaboration], 
\Journal{\PLB}{610}{2005}{199}, [hep-ex/0404002].

\bibitem{zeuslndstar}
S. Chekanov {\it et al.} [ZEUS Collaboration], 
\Journal{\PLB}{590}{2004}{143}, [hep-ex/0401017].

\bibitem{zeuslnjet}
S. Chekanov {\it et al.} [ZEUS Collaboration], 
                   \Journal{\NPB}{596}{2001}{3}, [hep-ex/0010019].

\bibitem{H1LNjets}
A. Aktas {\it et al.}  [H1 Collaboration], 
    \Journal{\EJC}{41}{2005}{273}, [hep-ex/0501074].
\end{thebibliography}
\end{document}